\documentclass[conference]{IEEEtran}
\IEEEoverridecommandlockouts
\usepackage{cite}
\usepackage{amsmath,amssymb,amsfonts}
\usepackage{newtxmath}
\usepackage{algorithmic}
\usepackage{graphicx}
\usepackage{textcomp}
\usepackage{xcolor}
\usepackage{enumitem}
\usepackage{caption}
\usepackage{subcaption}
\def\BibTeX{{\rm B\kern-.05em{\sc i\kern-.025em b}\kern-.08em
    T\kern-.1667em\lower.7ex\hbox{E}\kern-.125emX}}
\begin{document}
\title{Deutsch-Jozsa  and Bernstein-Vazirani algorithm using single‑particle discrete‑time 
quantum walk}
\author{\textit{}
    \IEEEauthorblockN{Ravi Sangwan\textsuperscript{1,2}, Vikas Ramaswamy\textsuperscript{2}, Henry Sukumar\textsuperscript{2} and Gudapati Naresh Raghava\textsuperscript{2}}  
    \IEEEauthorblockA{\textit{\textsuperscript{1}Department of Physics, Indian Institute of Space Science and Technology, Thiruvananthapuram, India}}  
    \IEEEauthorblockA{\textit{\textsuperscript{2}Quantum Technology Group, Centre for Development of Advanced Computing (C-DAC), Bangalore, India}  \\  
                    ravi.sc23m018@pg.iist.ac.in, vikasramaswamy@cdac.in, henrys@cdac.in, nareshraghava@cdac.in}    
}

\maketitle
\begin{abstract}
The paper introduces an efficient implementation of the Deutsch-Jozsa and Bernstein-Vazirani algorithm using the single‑particle discrete‑time quantum walk. We also provide a detailed optical framework with specific optical components to achieve these implementations in the photonic quantum walk scheme by simultaneously exploiting both polarization and path degrees of freedom. These implementations demonstrate improved resource efficiency while maintaining the exponential speedup characteristic of both algorithms. This work contributes to the growing field of universal quantum computing using single-particle discrete-time quantum walk.
\end{abstract}

\section{Introduction}
Quantum computing holds the promise of revolutionizing various fields by delivering exponential improvements in the efficiency of specific computational tasks compared to classical computing \cite{QCQI}. Among well-established quantum algorithms, the Deutsch-Jozsa algorithm \cite{DJ_Algorithm} and the Bernstein-Vazirani algorithm \cite{BV_Algo} are notable for their exponential speed-up over classical approaches. The Deutsch-Jozsa algorithm identifies whether a given function is constant or balanced using a single oracle query, and the Bernstein-Vazirani algorithm determines a binary string $s$ associated with a given function in a single oracle query. Several recent theoretical advancements have extended the Deutsch–Jozsa and Bernstein–Vazirani algorithms, including distributed versions \cite{Li2025}, generalizations beyond qubit systems\cite{Nagata2020}, and supersymmetric models \cite{Khemakhmia2025}. While these approaches improve scalability or structure, they rely on auxiliary qubits, entanglement, or complex operations, which limits their suitability for photonic platforms. In contrast, this paper presents an implementation of both algorithms using a single-particle discrete-time quantum walk, allowing the algorithm to be executed without the need for an auxiliary qubit, thereby improving resource efficiency by eliminating the additional quantum gates associated with it.

The following sections of this article provide a comprehensive exploration of the theoretical foundations and implementation details. Section~\ref{Universal quantum computing using single‑particle discrete‑time quantum walk} focuses on universal quantum computing using a single‑particle discrete‑time quantum walk along with a discussion on the use of photonic quantum walk for quantum computing. Section~\ref{Deutsch–Jozsa Algorithm} delves into the theoretical background of the Deutsch-Jozsa algorithm.  Section~\ref{Two-qubit Deutsch Jozsa} introduces the quantum walk-based approach to the Deutsch-Jozsa algorithm with and without using an auxiliary qubit along with a discussion on the implementation of both approaches in a photonic system. Section~\ref{Bernstein–Vazirani Algorithm} presents the theoretical background of the Bernstein–Vazirani algorithm and introduces the quantum walk-based scheme for its implementation. Section~\ref{Discussion and Analysis} presents a comparative analysis of the quantum walk schemes for the Bernstein-Vazirani and Deutsch-Jozsa algorithms, examining both implementations with and without an auxiliary qubit. Finally, Section~\ref{conclusion} provides a summary and concluding remarks on the study.
\section{Universal quantum computing using single‑particle discrete‑time quantum walk} \label{Universal quantum computing using single‑particle discrete‑time quantum walk}
Quantum Walks (QWs) are the quantum mechanical analogues of classical random walks influenced by the principles of quantum mechanics, leading to fundamentally different behavior and properties compared to their classical counterparts. QWs can be categorized into two types: discrete-time quantum walks (DTQWs) \cite{DTQW1} and continuous-time quantum walks (CTQWs) \cite{CTQW1}. DTQWs describe the evolution of a quantum walker in discrete time steps. In contrast, CTQWs describe the continuous evolution of a quantum state under a Hamiltonian that dictates its dynamical evolution. In the context of DTQWs, two main categories are distinguished: coin-based models and coin-less models \cite{Quantum_walks_review}. The coin-based models, are commonly referred to as discrete-time quantum walks (DTQWs). Coin-based quantum walks evolve under the influence of two key operators: the coin operator and the shift operator. The coin operator acts in the coin-Hilbert space $H_c$. Specifically, the coin operator for a $1D$ coin-based DTQW takes the form of a \(2 \times 2\) unitary matrix:
\begin{equation}\label{Eq:coin_operator}
\hat{C}( p, q, r, \theta) = e^{ip} \begin{bmatrix}
e^{iq} \cos(\theta) & e^{ir} \sin(\theta) \\
-e^{-ir} \sin(\theta) & e^{-iq} \cos(\theta)
\end{bmatrix},
\end{equation}
where \( p, q, r, \theta \in \mathbb{R} \). The shift operator acts in the position-Hilbert space \( H_s = \textit{span}\{|l\rangle\}, l \in \mathbb{Z} \), where \( \mathbb{Z} \) denotes the set of available position states. The conditional position shift operation governs the walker's movement on the basis of coin state, thereby influencing the evolution of the quantum walker in position space. The corresponding shift operators are defined as follows \cite{Qc_using_QW}: \begin{align}\label{Eq:Shift_operators}
 \hat{S}_{-}^a  &= \sum_{l \in \mathbb{Z}} \sum_{b} |a\rangle \langle a| \otimes |l-1\rangle \langle l| + |b \neq a\rangle \langle b| \otimes |l\rangle \langle l| \notag \\
 \hat{S}_{+}^b &= \sum_{l \in \mathbb{Z}} \sum_{a} |a \neq b\rangle \langle a| \otimes |l\rangle \langle l| + |b\rangle \langle b| \otimes |l+1\rangle \langle l|,
\end{align} where \( |a\rangle \) and \( |b\rangle \) represent basis states in the Hilbert space of the coin \( H_C \) \( \left(|a\rangle, |b\rangle \in \{ |0\rangle, |1\rangle \} \right)\).  The set of operators \( \{\hat{I}, \hat{S}_{+}^0, \hat{S}_{-}^0, \hat{S}_{+}^1, \hat{S}_{-}^1, \hat{C}_t\} \) forms a fundamental building block for constructing $1D$ coin-based DTQWs. Within the DTQWs framework these set of operators can be used to effectively realize universal set of quantum gates (Hadamard ($H$), Phase ($P$), and CNOT). Furthermore, the coin  and position states serve as a natural encoding scheme for qubits in a quantum system \cite{Qc_using_QW}.

The scheme proposed in~\cite{Exp_QC_using_QW} demonstrates the experimental realization of a universal set of quantum gates (Hadamard (H), Phase (P), and CNOT) with high fidelity using photonic quantum walks in combination with polarization and path degrees of freedom. In this approach, qubits can be encoded using both the path and polarization degrees of freedom of a photon. The path degree of freedom corresponds to the designated path (mode) number of the photon, while the polarization degree of freedom encodes additional information, typically as horizontal (\(H\)) or vertical (\(V\)) polarization states. The polarization Hilbert space serves a role analogous to the coin space in DTQWs, while the path degree of freedom corresponds to the position Hilbert space. One of the qubits is encoded in the polarization degrees of freedom, while additional qubits are encoded in the path degrees of freedom, and the implementation of quantum gates is achieved through a combination of a path (mode)-dependent quantum coin operation, typically realized using a sequence of QWP-HWP-QWP (quarter-wave plate, half-wave plate, quarter-wave plate) or other waveplate configurations, along with a conditioned shift operator, which is facilitated by polarizing beam splitters (PBS) or beam splitters (BS). 

\section{Deutsch-Jozsa Algorithm}\label{Deutsch–Jozsa Algorithm} 
The Deutsch-Jozsa algorithm signifies a substantial development in quantum computing, demonstrating an exponential speedup over classical methods for identifying whether the input function is constant or balanced \cite{DJ_Algorithm}. The algorithm takes a function \( f: \{0,1\}^n \to \{0,1\} \) as input, where each \( n \)-bit input string is mapped to a binary output. The function must satisfy one of two conditions: it is either constant, giving the same output for all inputs, or balanced, giving $0$ for half of the inputs and $1$ for the other half. The objective is to identify whether the function is constant or balanced while minimizing the number of function evaluations. Classically, in the worst-case scenario, at least (\( 2^{n-1} + 1 \)) function evaluations are required. However, the Deutsch–Jozsa algorithm achieves this with a single function evaluation. For a function with an input string of $n$-bits, the Deutsch-Jozsa algorithm can be implemented using $n+1$ qubits (standard approach with auxiliary qubit) or $n$ qubits (without auxiliary qubit). The standard approach with an auxiliary qubit follows the conventional oracle design, and the auxiliary qubit encodes the function's output in a phase. The $n$-qubit approach (without an auxiliary qubit) is more resource-efficient, as it eliminates the need for an extra auxiliary qubit, but requires a more intricate oracle construction that directly applies the phase transformation. The procedure for both approaches, the standard approach (with an auxiliary qubit) and the approach without an auxiliary qubit, is outlined below:  
\begin{enumerate}
    \item Standard approach (with auxiliary qubit):
    \begin{enumerate}[label=\arabic{enumi}.\arabic*)]
     \item The quantum system is prepared in an $n+1$ qubit state, where $n$ qubits represent the input register and an additional auxiliary qubit serves as the function output register: \begin{equation}\label{eq:dj_initial state_with_auxiliary)}
       | \phi_0 \rangle = |0\rangle^{\otimes n} \otimes |1\rangle
    \end{equation}
    \item  Each qubit undergoes a hadamard transformation: \begin{align} \label{eq:dj_superposiotion_state_with_auxiliary)} | \phi_1 \rangle &= H^{\otimes (n+1)} | \psi_0 \rangle \notag\\ &= \left(\frac{1}{\sqrt{2^n}} \sum_{x =0}^{2^n-1} |x\rangle \right) \otimes \frac{1}{\sqrt{2}} (|0\rangle - |1\rangle)
    \end{align}       
    \item The function \( f \) is encoded into a quantum oracle, which applies a phase shift conditioned on the output of the function. \begin{equation}\label{eq:dj_oracle_with_auxiliary)}
        U_f |x\rangle |y\rangle = |x\rangle |y \oplus f(x)\rangle
        \end{equation} As the auxiliary qubit is in the state (\( |0\rangle - |1\rangle \)), the oracle introduces a phase shift of $(-1)^{f(x)}$:
\begin{equation}\label{eq:dj_superposiotion_state_with_auxiliary_1}
U_f | \phi_1 \rangle = \frac{1}{\sqrt{2^n}} \sum_{x =0}^{2^n-1} (-1)^{f(x)} |x\rangle \otimes \frac{1}{\sqrt{2}} (|0\rangle - |1\rangle)
\end{equation}
    \item A second Hadamard transformation is applied to the input register \begin{align}\label{eq:psi2_state_with_auxiliary}
       | \phi_2 \rangle &=  H^{\otimes n} | \phi_1 \rangle \notag\\ &=  \frac{1}{2^n} \sum_{z =0}^{2^n-1} \left(\sum_{x =0}^{2^n-1} (-1)^{x \cdot z + f(x)} \right) |z\rangle \otimes \frac{1}{\sqrt{2}} (|0\rangle - |1\rangle)
    \end{align} From \eqref{eq:psi2_state_with_auxiliary} the probability of measuring $z = 0$ , coresponding to $|0\rangle^{\otimes n}$, is $$ \left| \frac{1}{2^{n}} \sum_{x=0}^{2^{n}-1} (-1)^{f(x)} \right|^{2}
$$  which is equal to $1$ if $f(x)$ is constant and $0$ if $f(x)$ is balanced.
\item Measurement of the first \( n \) qubits yields \( |0\rangle^{\otimes n} \) with probability $1$ if \( f(x) \) is constant; if \( f(x) \) is balanced, the probability is $0$.\end{enumerate}
    \item Approach without using auxiliary qubit:
    \begin{enumerate}[label=\arabic{enumi}.\arabic*)]
    \item  The quantum system is prepared in an \( n\) qubit state: \begin{equation}\label{eq:dj_initial state_without_auxiliary)}
       |\psi_0 \rangle = |0\rangle^{\otimes n}
    \end{equation}
    \item  Each qubit undergoes a Hadamard transformation: \begin{align}\label{eq:dj_superposiotion_state_without_auxiliary)}
       | \psi_1 \rangle &= H^{\otimes n} | \psi_0 \rangle \notag\\ &= \left(\frac{1}{\sqrt{2^n}} \sum_{x =0}^{2^n-1} |x\rangle \right)
    \end{align}       
    \item  The function \( f \) is encoded into a quantum oracle, which applies a phase shift conditioned on the function's output. \begin{equation}\label{eq:dj_oracle_without_auxiliary)}
        U_f |x\rangle  = (-1)^{f(x)}|x\rangle
    \end{equation} So,
\begin{equation}\label{eq:dj_superposiotion_state_without_auxiliary_1}
U_f | \psi_1 \rangle = \frac{1}{\sqrt{2^n}} \sum_{x=0}^{2^{n}-1} (-1)^{f(x)} |x\rangle
\end{equation}
    \item  Hadamard gates are applied to all qubits: \begin{align}\label{eq:psi2_state_without_auxiliary}
       | \psi_2 \rangle &=  H^{\otimes n} | \psi_1 \rangle \notag\\ &=  \frac{1}{2^n} \sum_{z=0}^{2^{n}-1} \left( \sum_{x=0}^{2^{n}-1} (-1)^{x \cdot z + f(x)} \right) |z\rangle 
    \end{align} From \eqref{eq:psi2_state_without_auxiliary} the probability of measuring $z = 0$ , coresponding to $|0\rangle^{\otimes n}$, is $$ \left| \frac{1}{2^{n}} \sum_{x=0}^{2^{n}-1} (-1)^{f(x)} \right|^{2}$$  which is equal to $1$ if $f(x)$ is constant and $0$ if $f(x)$ is balanced.
\item Measurement of qubits yields \( |0\rangle^{\otimes n} \) with probability $1$ if \( f(x) \) is constant; if \( f(x) \) is balanced, the probability is $0$.\end{enumerate} \end{enumerate} \section{Two-qubit Deutsch Jozsa}\label{Two-qubit Deutsch Jozsa}
\begin{figure}[htbt]
    \centering
    \renewcommand{\thesubfigure}{\roman{subfigure}}
    \begin{subfigure}[b]{0.107\textwidth}
        \centerline{\includegraphics[width=\textwidth]{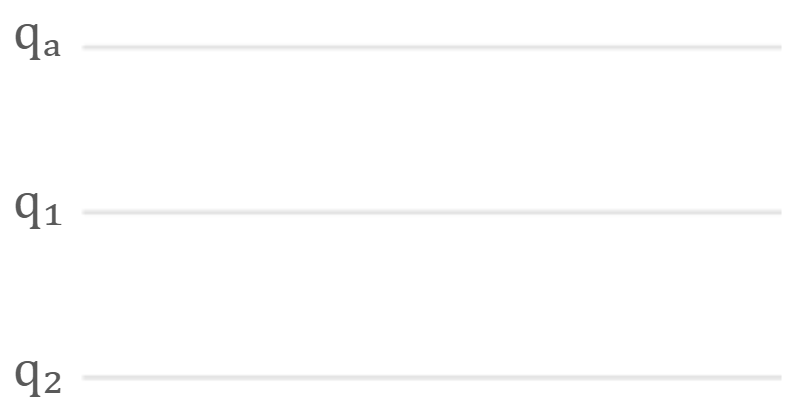}}
         \caption{}
    \end{subfigure}
    \hfill
    \begin{subfigure}[b]{0.1\textwidth}
        \centerline{\includegraphics[width=\textwidth]{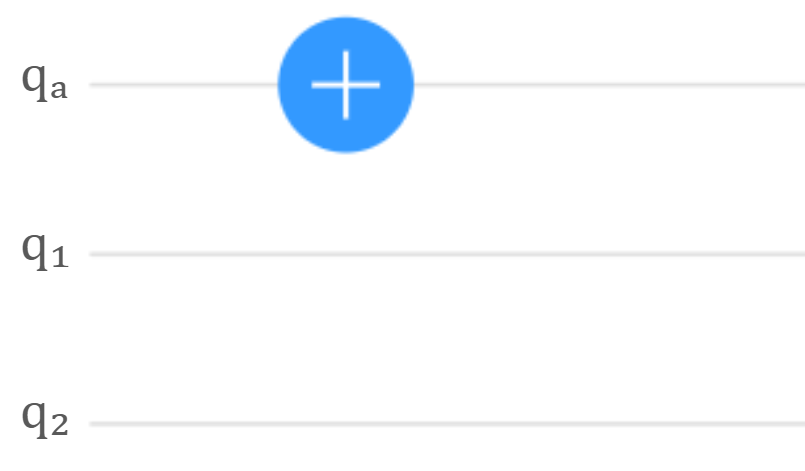}}
         \caption{}
    \end{subfigure}
    \hfill
    \begin{subfigure}[b]{0.097\textwidth}
        \centerline{\includegraphics[width=\textwidth]{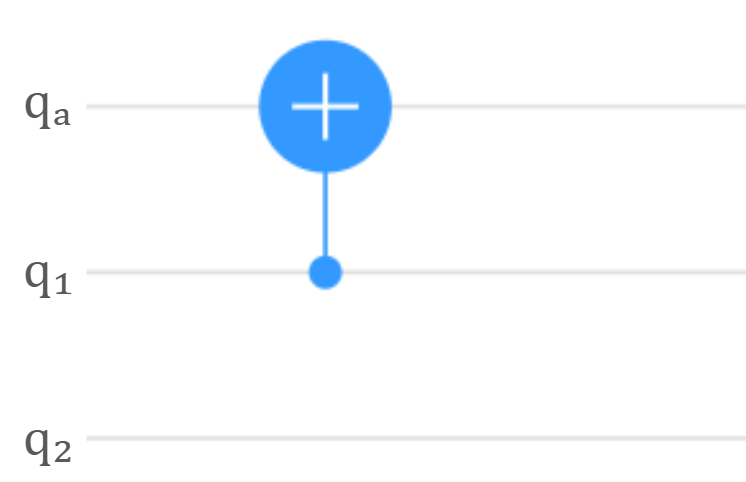}}
         \caption{}
    \end{subfigure}
    \hfill
    \begin{subfigure}[b]{0.098\textwidth}
        \centerline{\includegraphics[width=\textwidth]{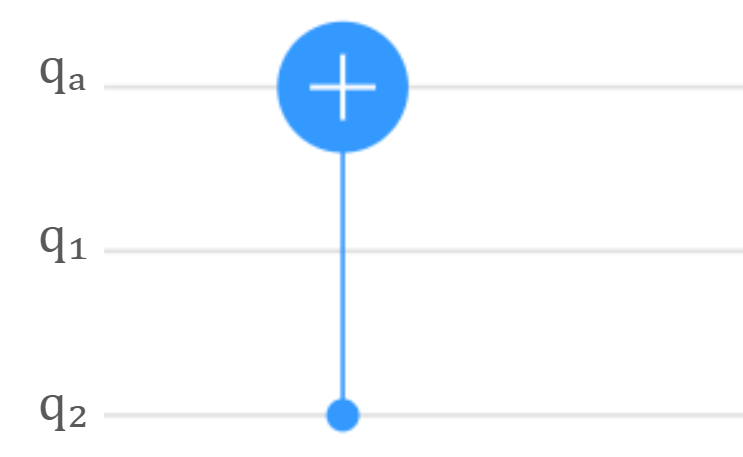}}
         \caption{}
    \end{subfigure}
       \hfill
    \begin{subfigure}[b]{0.107\textwidth}
        \centerline{\includegraphics[width=\textwidth]{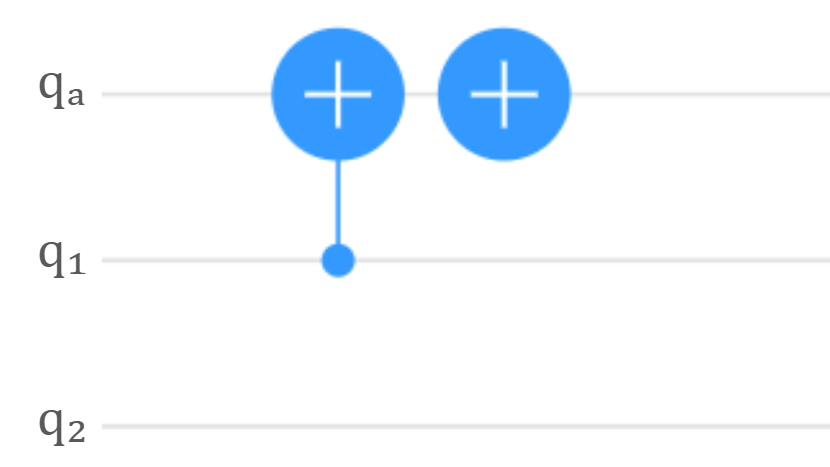}}
         \caption{}
    \end{subfigure}  
    \hfill
    \begin{subfigure}[b]{0.1\textwidth}
       \centerline{\includegraphics[width=\textwidth]{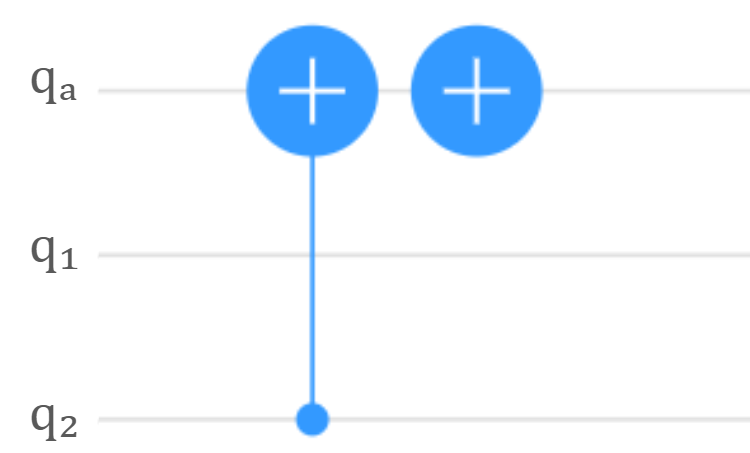}}
         \caption{}
    \end{subfigure}
    \hfill
    \begin{subfigure}[b]{0.097\textwidth}
        \centerline{\includegraphics[width=\textwidth]{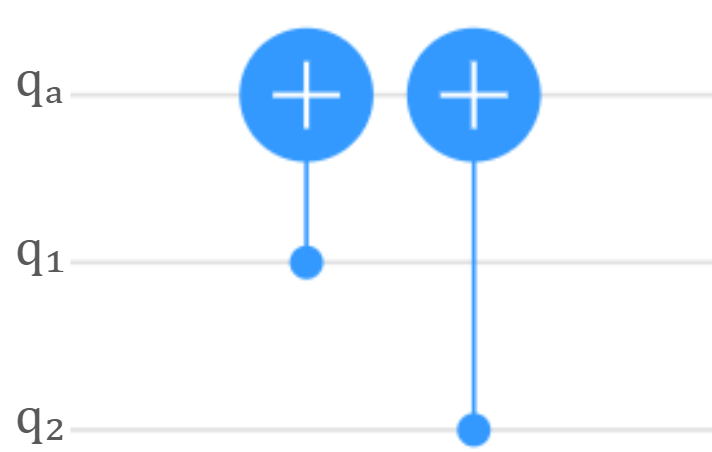}}
         \caption{}
    \end{subfigure}
    \hfill
    \begin{subfigure}[b]{0.108\textwidth}
        \centerline{\includegraphics[width=\textwidth]{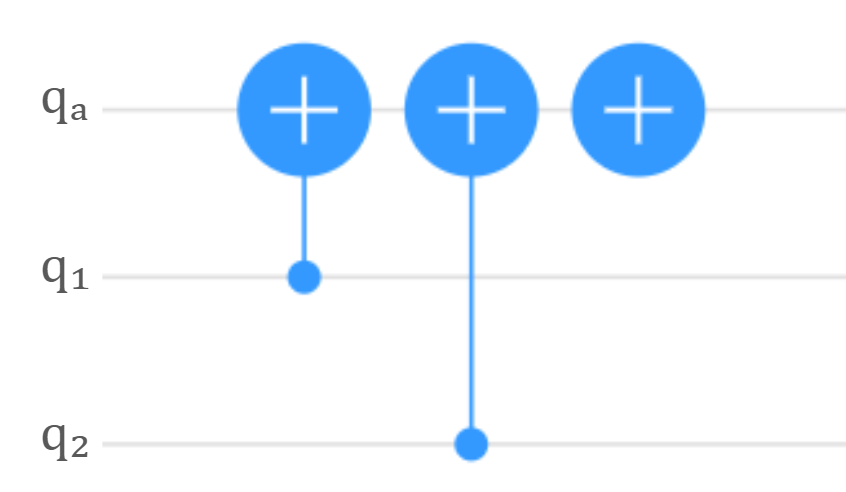}}
         \caption{}
    \end{subfigure}
    \caption{ Oracle implementations for all two-bit Boolean functions  that are either constant or balanced given in Table~\ref{tab:two_bit_functions}, by using $X$ and CNOT gates in standard circuit model with auxiliary qubit~\cite{Qniverse}. The $q_a$, $q_1$, and $q_2$ represents auxiliary, first working and second working  qubit, respectively.}
    \label{fig:dj_two_qubit_circuit}
\end{figure}

In this section, we present the quantum walk scheme for two-qubit Deutsch-Jozsa algorithm with auxiliary qubit and without using auxiliary qubit along with  a discussion on the implementation of both schemes in a photonic system. For the two-bit case, the function $f(x_1, x_2)$ has eight possible realizations as shown in Table~\ref{tab:two_bit_functions}.
In the standard circuit model the oracle~\eqref{eq:dj_oracle_with_auxiliary)} is achieved utilizing a sequence of $X$ and CNOT gates applied to the auxiliary qubit. The choice and configuration of these gates depend on the specific function being implemented, ensuring that the oracle correctly encodes the function's behavior. The detailed oracle implementations for each function are presented in  Fig.~\ref{fig:dj_two_qubit_circuit}. 
\subsection{Quantum walk scheme using auxiliary qubit}\label{Quantum walk scheme using auxiliary qubit} 
\begin{figure}[htbp]
    \centering
    \renewcommand{\thesubfigure}{\roman{subfigure}}
    \begin{subfigure}[b]{0.108\textwidth}
        \centerline{\includegraphics[width=\textwidth]{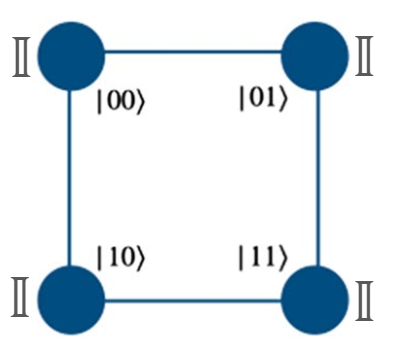}}
         \caption{}
    \end{subfigure}
    \hfill
    \begin{subfigure}[b]{0.12\textwidth}
        \centerline{\includegraphics[width=\textwidth]{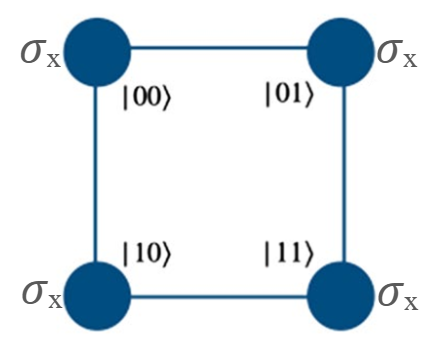}}
         \caption{}
    \end{subfigure}
    \hfill
    \begin{subfigure}[b]{0.12\textwidth}
        \centerline{\includegraphics[width=\textwidth]{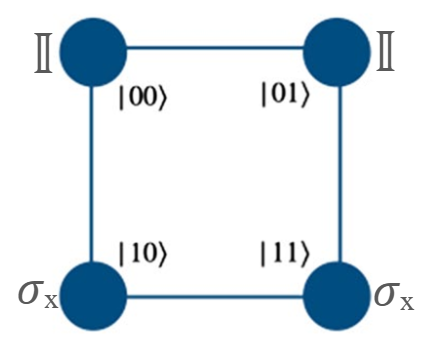}}
         \caption{}
    \end{subfigure}
    \hfill
    \begin{subfigure}[b]{0.117\textwidth}
       \centerline{\includegraphics[width=\textwidth]{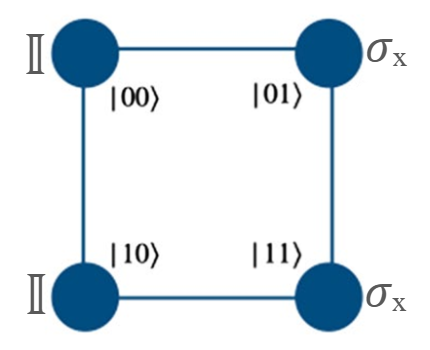}}
         \caption{}
    \end{subfigure}
       \hfill
    \begin{subfigure}[b]{0.117\textwidth}
        \centerline{\includegraphics[width=\textwidth]{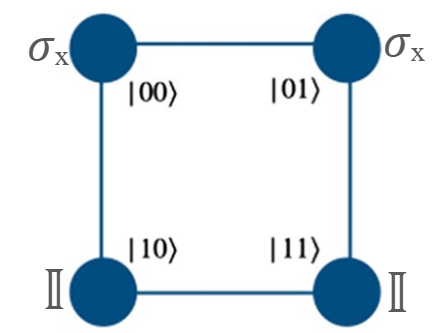}}
         \caption{}
    \end{subfigure}   
    \hfill
    \begin{subfigure}[b]{0.116\textwidth}
        \centerline{\includegraphics[width=\textwidth]{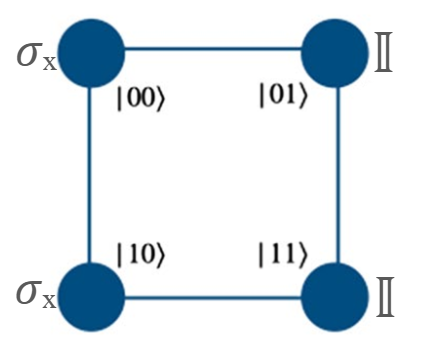}}
         \caption{}
    \end{subfigure}
    \hfill
    \begin{subfigure}[b]{0.117\textwidth}
        \centerline{\includegraphics[width=\textwidth]{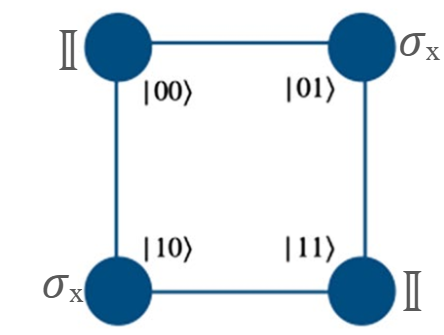}}
         \caption{}
    \end{subfigure}
    \hfill
    \begin{subfigure}[b]{0.12\textwidth}
        \centerline{\includegraphics[width=\textwidth]{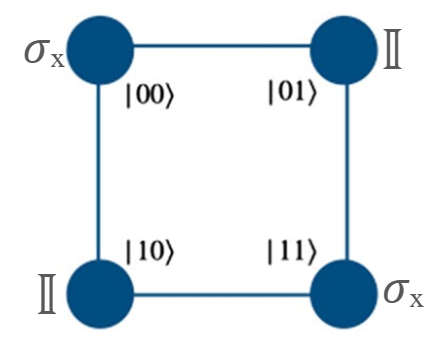}}
         \caption{}
    \end{subfigure}
    \caption{Schematic illustration of the quantum walk based oracle implementations with auxiliary qubit for all two-bit boolean functions that are either constant or balanced given in Table~\ref{tab:two_bit_functions}. Here, $\mathbb{I}$ represents the identity coin operator together with an identity shift operator identity shift operator ($\hat{C}( 0, 0, 0, 0) \otimes \hat{I}$). }
    \label{fig:dj_oracle_two_qubit_qw}
\end{figure}

\begin{figure}[htbt]
    \centering
    \renewcommand{\thesubfigure}{\roman{subfigure}}
    \begin{subfigure}[b]{0.11\textwidth}
        \centerline{\includegraphics[width=\textwidth]{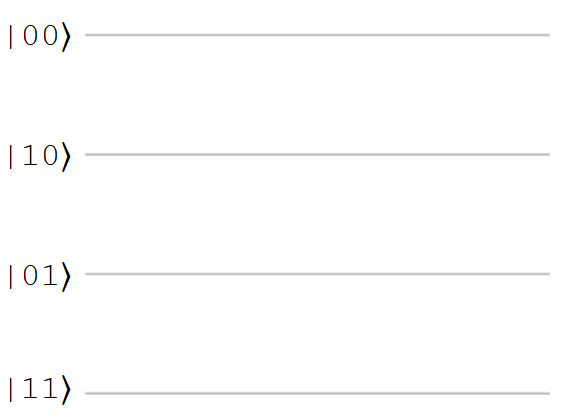}}
         \caption{}
    \end{subfigure}
    \hfill
    \begin{subfigure}[b]{0.1\textwidth}
        \centerline{\includegraphics[width=\textwidth]{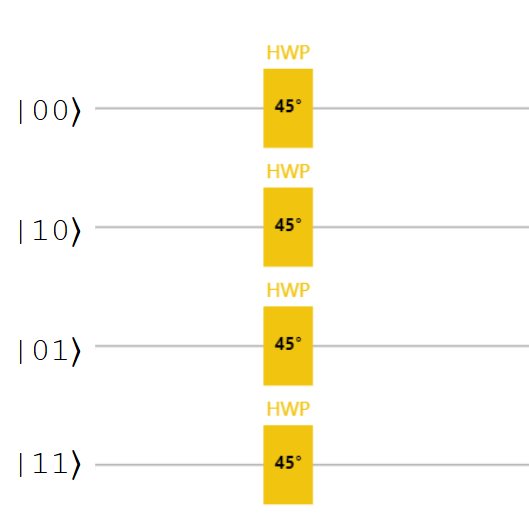}}
         \caption{}
    \end{subfigure}
    \hfill
    \begin{subfigure}[b]{0.087\textwidth}
        \centerline{\includegraphics[width=\textwidth]{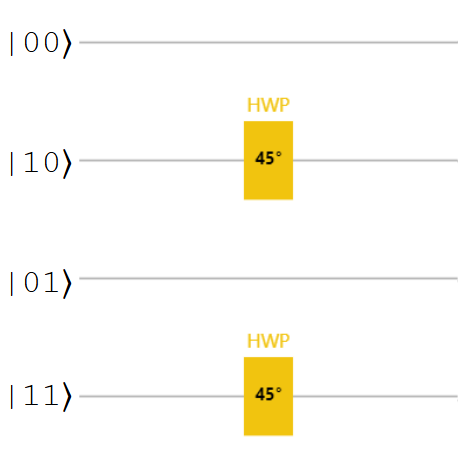}}
         \caption{}
    \end{subfigure}
    \hfill
    \begin{subfigure}[b]{0.095\textwidth}
       \centerline{\includegraphics[width=\textwidth]{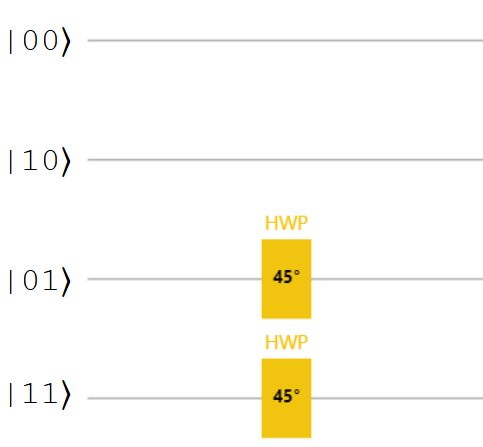}}
         \caption{}
    \end{subfigure}
       \hfill
    \begin{subfigure}[b]{0.093\textwidth}
        \centerline{\includegraphics[width=\textwidth]{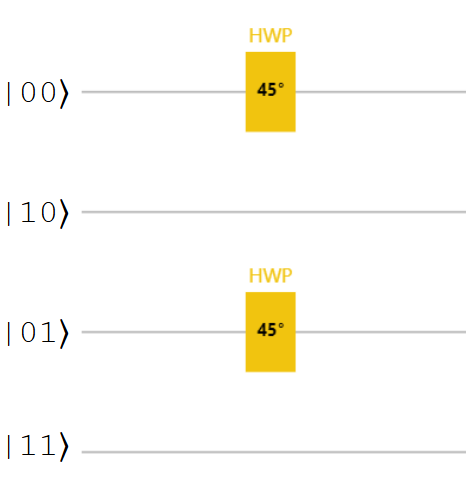}}
         \caption{}
    \end{subfigure}   
    \hfill
    \begin{subfigure}[b]{0.093\textwidth}
        \centerline{\includegraphics[width=\textwidth]{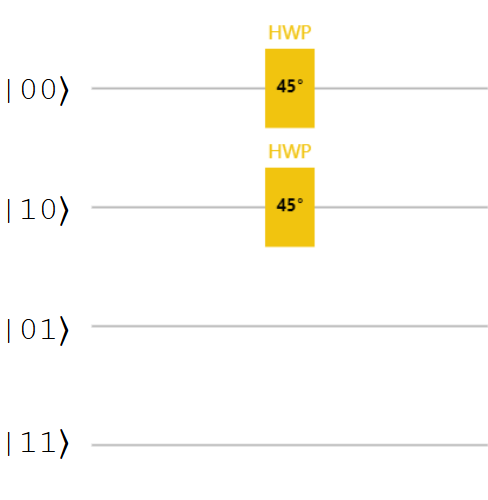}}
         \caption{}
    \end{subfigure}
    \hfill
    \begin{subfigure}[b]{0.099\textwidth}
        \centerline{\includegraphics[width=\textwidth]{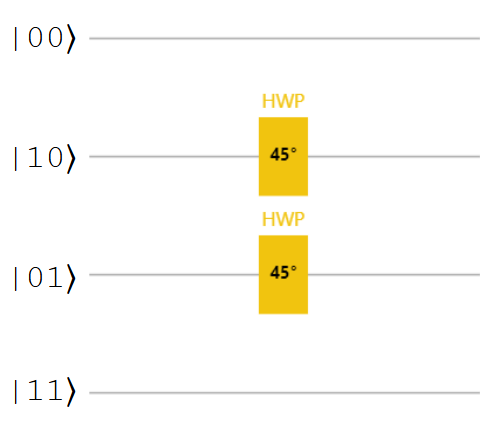}}
         \caption{}
    \end{subfigure}
    \hfill
    \begin{subfigure}[b]{0.09\textwidth}
        \centerline{\includegraphics[width=\textwidth]{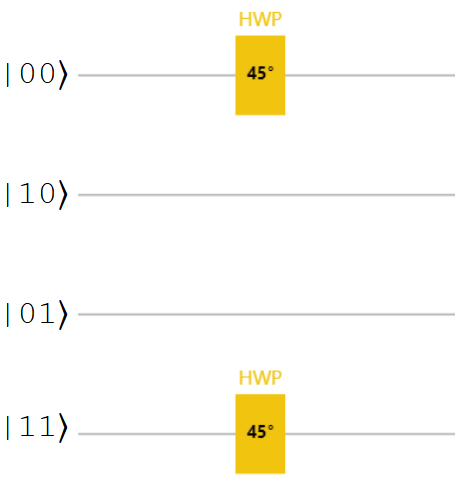}}
         \caption{}
    \end{subfigure}
    \caption{Schematic illustration of the photonic implementation of the operations shown in Fig.~\ref{fig:dj_oracle_two_qubit_qw}. Here, the four path degrees ($|00\rangle,|10\rangle,|01\rangle, \text{and}~|11\rangle$) of freedom represent two working qubits and the polarization state of the photon repersents the auxiliary qubit.}
    \label{fig:dj_oracle_two_qubit_qw_onps}
\end{figure}

The quantum system is initialized in a three-qubit state i.e., the particle undergoes a quantum walk on an closed graph  consisting of four vertices where the coin state represents the auxiliary qubit  and position space \textit{span}\{$|00\rangle, |01\rangle, |11\rangle, |10\rangle$\} \begin{table}[htbt]
    \caption{Inputs and outputs of all possible two-bit Boolean functions that are either constant or balanced. Functions (\textit{i}) and (\textit{ii}) correspond to constant functions, while functions (\textit{iii}) to (\textit{viii}) represent balanced functions. Specifically, 
$f(x_1,x_2)$ is defined as: (\textit{i}) $0$, (\textit{ii})$ 1$ , (\textit{iii}) $x_1$, (\textit{iv}) $x_2$, (\textit{v}) $\overline{x_1}$, (\textit{vi}) $\overline{x_2}$, (\textit{vii}) $ x_1 \oplus x_2$, and (\textit{viii}) $\overline{x_1 \oplus x_2}$.}
\begin{center}
\begin{tabular}{|c|c|c|c|c|c|c|c|c|}
        \hline
        \multicolumn{1}{|c|}{\textbf{Inputs}} & \multicolumn{8}{c|}{\textbf{Functions $f(x_1,x_2)$}} \\  
        \hline
        \textbf{$(x_1,x_2)$} &  (\textit{i})  & (\textit{ii})  & (\textit{iii})  & (\textit{iv})  & (\textit{v}) & (\textit{vi})  & (\textit{vii}) & (\textit{viii})   \\
        \hline
        \( (0,0) \) & 0 & 1 & 0 & 0 & 1 & 1 & 0 & 1 \\
        \( (0,1) \) & 0 & 1 & 0 & 1 & 1 & 0 & 1 & 0 \\
        \( (1,0) \) & 0 & 1 & 1 & 0 & 0 & 1 & 1 & 0 \\
        \( (1,1) \) & 0 & 1 & 1 & 1 & 0 & 0 & 0 & 1 \\
        \hline
    \end{tabular}
\label{tab:two_bit_functions}
\end{center}
\end{table}
represents two working qubits.   Initially, the particle starts  with  $|0\rangle$ coin state and $|00\rangle$ position state, after that coin operation $\hat{C}( -\pi/2, 0, \pi/2, \pi/2)$~\eqref{Eq:coin_operator} is applied on coin (particle) state followed by the identity operation on position space to obtain the coin state $|1\rangle$. Both Hadamard transformations, as outlined in~\eqref{eq:dj_superposiotion_state_with_auxiliary)} and~\eqref{eq:psi2_state_with_auxiliary}, are realized through a sequence of shift and coin operations. Specifically, the Hadamard gate on the auxiliary qubit (coin state) is implemented via the coin operation $\hat{C}(-\pi/2, -\pi/2, \pi/2, \pi/4)$~\eqref{Eq:coin_operator} applied to the coin (particle) state, followed by an identity operation on the position space. The Hadamard gates acting on the working qubits in position space are implemented by evolving the coin state into a superposition across the position space. This is achieved through appropriate combinations of Pauli-\(X\), \(Z\), and Hadamard operations on the coin space, along with shift operations $\{\hat{S}_{-}^a, \hat{S}_{-}^b\}$~\eqref{Eq:Shift_operators}. The detailed mathematical expressions for these operations can be found in~\cite{Qc_using_QW}. The quantum walk based oracle implementations for all two-bit boolean functions that are either constant or balanced with auxiliary qubit~\eqref{eq:dj_oracle_with_auxiliary)} can be achieved by using a position-dependent coin operation $\hat{C}( -\pi/2, 0, \pi/2, \pi/2)$ together with an identity shift operator i.e., $\sigma_\text{x} = \hat{C}(  -\pi/2, 0, \pi/2, \pi/2) \otimes \hat{I}$ as shown in Fig~\ref{fig:dj_oracle_two_qubit_qw}. The operations for the oracle implementation shown in Fig.~\ref{fig:dj_oracle_two_qubit_qw} can be implemented in the photonic system using HWPs at angle $\pi/4$ as shown in Fig.~\ref{fig:dj_oracle_two_qubit_qw_onps}. The corresponding Jones matrices for the HWP rotated by angle $\alpha$ w.r.t horizontal axis is given by  \begin{equation}\label{Eq:HWP_JOnes_Matrix}
\text{HWP}(\alpha) = 
\begin{pmatrix}
\cos{2\alpha} & \sin{2\alpha} \\
\sin{2\alpha} & -\cos{2\alpha}
\end{pmatrix}.
\end{equation} 

 

\begin{figure}[htbt]
    \centering
    \includegraphics[width=1\linewidth]{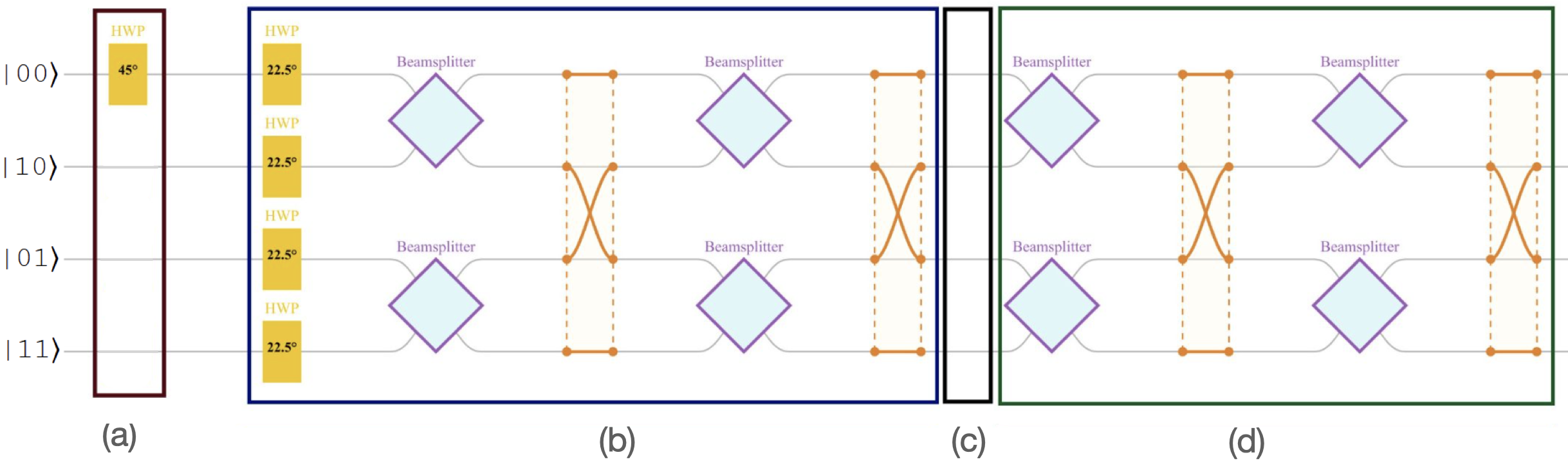}
    \caption{The photonic circuit for the photonic quantum walk based two-qubit Deutsch-Jozsa algorithm using auxiliary qubit for the function ($f(x_1,x_2)=0$). The four path degrees ($|00\rangle,|01\rangle,|11\rangle, \text{and}~|10\rangle$) of freedom represent two working qubits and the polarization state represents the auxiliary qubit. Here, (a), (b), (c), and (d) boxes represent $X$ gate on polarization state (auxiliary qubit), $H$ gate on all the qubits, oracle, and $H$ gate on two working qubits, respectively.}
    \label{fig:2dj_aux_ps_full}
\end{figure}

In the implementation of this scheme using photonic quantum walk, a photon undergoes a quantum walk within a four-path (mode) photonic circuit where the polarization state of the photon  represents the auxiliary qubit and four path degrees of freedom represent two working qubits. The photon starts with  $|H\rangle$ polarization state in the $1$st path ($|00\rangle$),  after that HWP($\pi/4$) as in \eqref{Eq:HWP_JOnes_Matrix} is applied on the $1$st path ($|00\rangle$) to get $|V\rangle$ polarization state.   
Both Hadamard transformations, as described in~\eqref{eq:dj_superposiotion_state_with_auxiliary)} and~\eqref{eq:psi2_state_with_auxiliary}, are implemented using half-wave plates (HWPs) set at $\pi/8$ (HWP($\pi/8$))~\eqref{Eq:HWP_JOnes_Matrix}, 50:50 beam splitters (BSs), and mode permuters. The mode permuter facilitates the manipulation and reordering of photon paths, while the 50:50 beam splitter applies the Hadamard transformation to the photon's path degree of freedom. The transformation matrix for a 50:50 beam splitter, essential for generating superposition states, is given by  
\begin{equation}\label{Eq:BS_matrix}
   BS = 
\frac{1}{\sqrt{2}}
\begin{bmatrix} 1 & 1 \\ 1 & -1  \end{bmatrix}. 
\end{equation}  The Hadamard transformations, as described in~\eqref{eq:dj_superposiotion_state_with_auxiliary)} and~\eqref{eq:psi2_state_with_auxiliary}, are highlighted in (b) and (d) boxes in Fig.~\ref{fig:2dj_aux_ps_full}. The sequence of operations for the implementation of oracles are depicted in Fig.~\ref{fig:dj_oracle_two_qubit_qw_onps}.  The photonic circuit for the two-qubit photonic quantum walk-based  Deutsch-Jozsa algorithm using an auxiliary qubit for the function $f(x_1,x_2) = 0$ is shown in Fig.~\ref{fig:2dj_aux_ps_full}, where the (c) box represents the oracle. For the other functions listed in Table~\ref{tab:two_bit_functions}, this box is replaced with the corresponding photonic circuit that implements the oracle for the given function, as shown in Fig.~\ref{fig:dj_oracle_two_qubit_qw_onps}. 

\subsection{Quantum walk scheme without using auxiliary qubit}\label{Quantum walk scheme without using auxiliary qubit} 
\begin{figure}[htbt]
    \centering
    \renewcommand{\thesubfigure}{\roman{subfigure}}
    \begin{subfigure}[b]{0.035\textwidth}
        \centerline{\includegraphics[width=\textwidth]{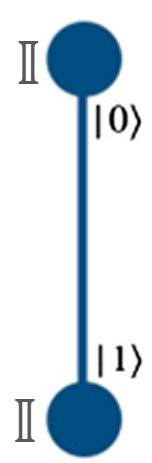}}
         \caption{}
    \end{subfigure}
    \hfill
    \begin{subfigure}[b]{0.043\textwidth}
        \centerline{\includegraphics[width=\textwidth]{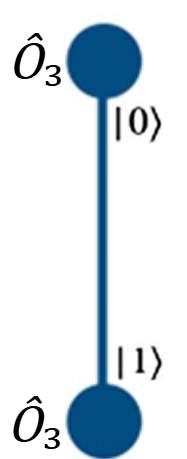}}
         \caption{}
    \end{subfigure}
    \hfill
    \begin{subfigure}[b]{0.04\textwidth}
        \centerline{\includegraphics[width=\textwidth]{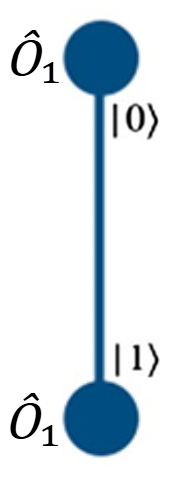}}
         \caption{}
    \end{subfigure}
    \hfill
    \begin{subfigure}[b]{0.0385\textwidth}
        \centerline{\includegraphics[width=\textwidth]{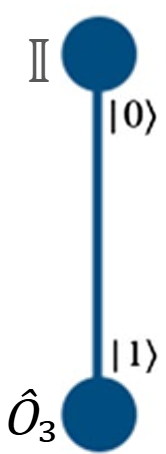}}
         \caption{}
    \end{subfigure}
    \hfill
       \begin{subfigure}[b]{0.044\textwidth}
       \centerline{\includegraphics[width=\textwidth]{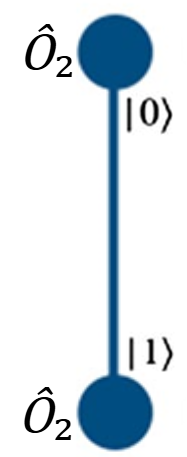}}
         \caption{}
    \end{subfigure}
    \hfill
    \begin{subfigure}[b]{0.040\textwidth}
        \centerline{\includegraphics[width=\textwidth]{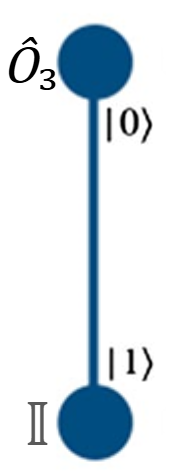}}
         \caption{}
    \end{subfigure}
    \begin{subfigure}[b]{0.059\textwidth}
        \centerline{\includegraphics[width=\textwidth]{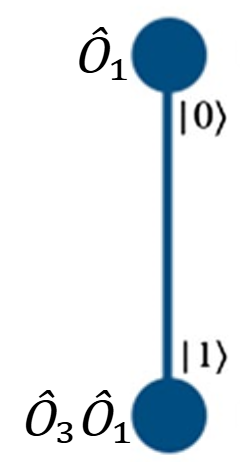}}
         \caption{}
    \end{subfigure}
    \hfill
    \begin{subfigure}[b]{0.055\textwidth}
        \centerline{\includegraphics[width=\textwidth]{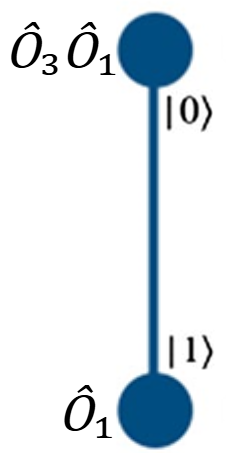}}
         \caption{}
    \end{subfigure}
    \caption{Schematic illustration of the  quantum walk based oracle implementations without auxiliary qubit for all two-bit boolean functions that are either constant or balanced given in Table~\ref{tab:two_bit_functions}, by using identity operation $\mathbb{I}$ and  three position dependent evolution operators ($\hat{O}_1, \hat{O}_2,\hat{O}_3$) as defined in~\eqref{eq:operators_twoqubit_dj_usingauxiliary_qw}.}
    \label{fig:dj_oracle_two_qubit_qw_without_aux}
\end{figure}

\begin{figure}[htbt]
    \centering
    \renewcommand{\thesubfigure}{\roman{subfigure}}
    \begin{subfigure}[b]{0.079\textwidth}
        \centerline{\includegraphics[width=\textwidth]{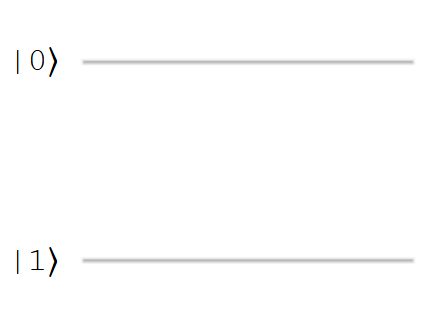}}
         \caption{}
    \end{subfigure}
    \hfill
    \begin{subfigure}[b]{0.11\textwidth}
        \centerline{\includegraphics[width=\textwidth]{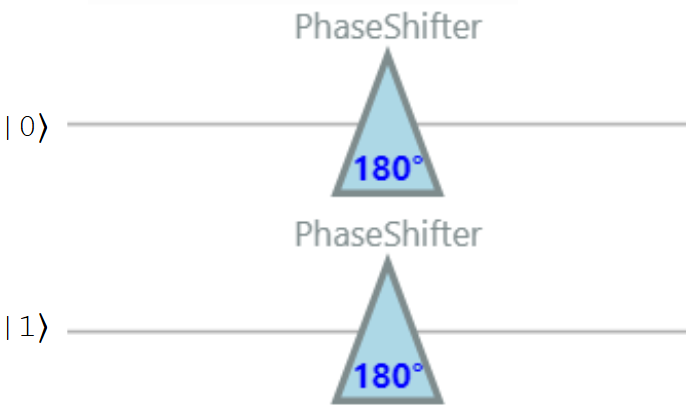}}
         \caption{}
    \end{subfigure}
    \hfill
    \begin{subfigure}[b]{0.11\textwidth}
        \centerline{\includegraphics[width=\textwidth]{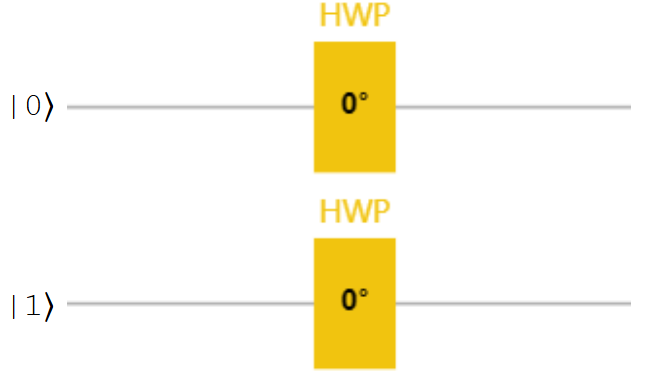}}
         \caption{}
    \end{subfigure}
    \hfill
    \begin{subfigure}[b]{0.11\textwidth}
       \centerline{\includegraphics[width=\textwidth]{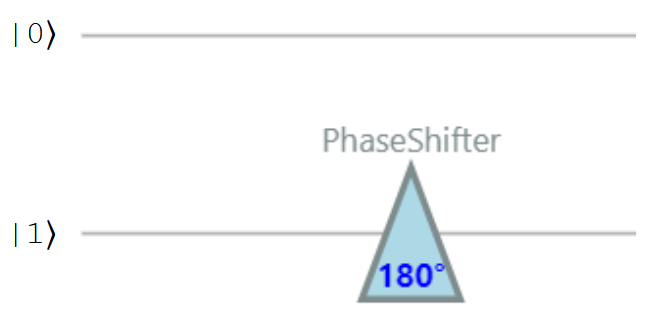}}
         \caption{}
    \end{subfigure}
       \hfill
    \begin{subfigure}[b]{0.09\textwidth}
        \centerline{\includegraphics[width=\textwidth]{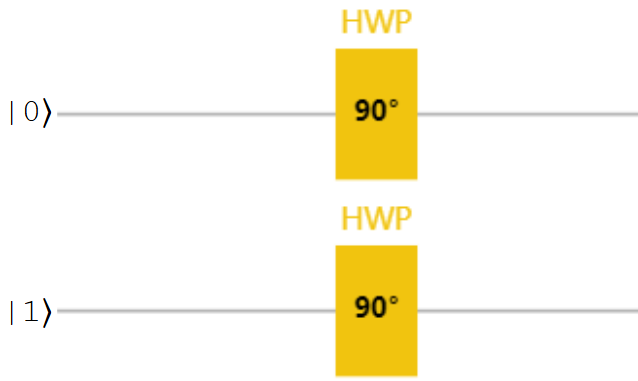}}
         \caption{}
    \end{subfigure}
    \hfill
    \begin{subfigure}[b]{0.107\textwidth}
        \centerline{\includegraphics[width=\textwidth]{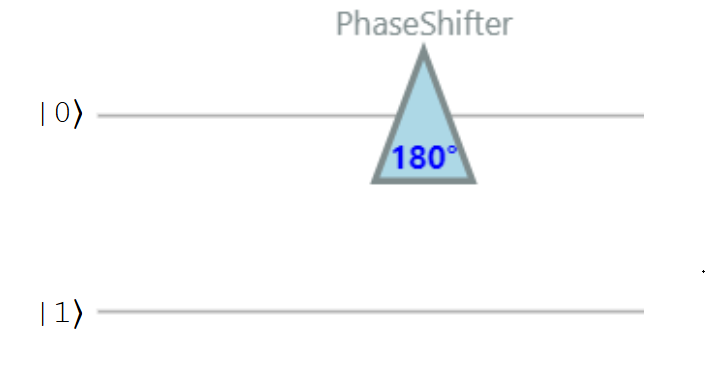}}
         \caption{}
    \end{subfigure}
    \hfill
    \begin{subfigure}[b]{0.123\textwidth}
        \centerline{\includegraphics[width=\textwidth]{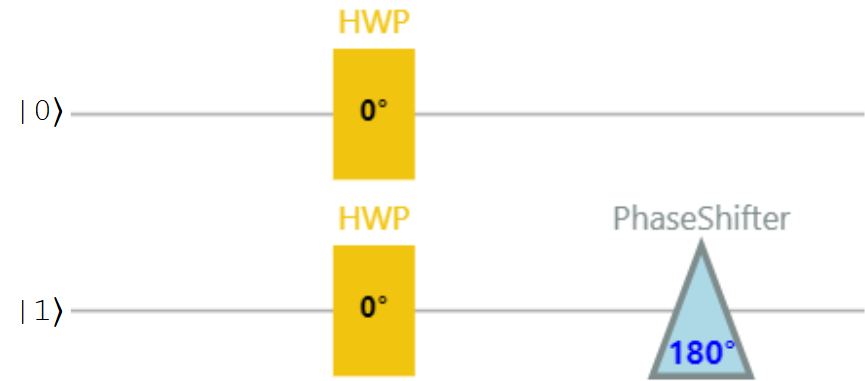}}
         \caption{}
    \end{subfigure}
    \hfill
    \begin{subfigure}[b]{0.115\textwidth}
        \centerline{\includegraphics[width=\textwidth]{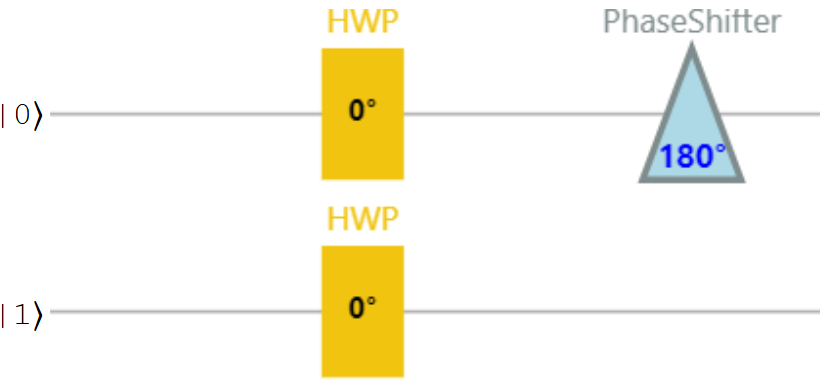}}
         \caption{}
    \end{subfigure}
    \caption{Schematic illustration of the photonic implementation of the operations shown in Fig.~\ref{fig:dj_oracle_two_qubit_qw_without_aux}. Here, the two path degrees ($|0\rangle,|1\rangle$) of freedom represent second qubit and the polarization state of the photon represents the first qubit.}
    \label{fig:dj_oracle_two_qubit_qw_without_aux_on_ps}
\end{figure}

The quantum system is initialized in a two-qubit state i.e., the particle undergoes a quantum walk on an open graph consisting of two vertices where the coin state represents the first qubit and position space \textit{span}\{$|0\rangle, |1\rangle$\} represents the second qubit. Initially, the particle starts with $|0\rangle$ coin state and $|0\rangle$ position state. Both Hadamard transformations, as described in~\eqref{eq:dj_superposiotion_state_without_auxiliary_1} and~\eqref{eq:psi2_state_without_auxiliary}, are implemented through a sequence of shift and coin operations as discussed in Section~\ref{Quantum walk scheme using auxiliary qubit}. The quantum walk based oracle implementations for all two-bit boolean functions that are either constant or balanced without auxiliary qubit~\eqref{eq:dj_oracle_without_auxiliary)} can be achieved  by applying identity operation $\mathbb{I}$ and  three position dependent evolution operators ($\hat{O}_1, \hat{O}_2,\hat{O}_3$): \begin{align}
\hat{O}_1 &= \hat{C}\left( \tfrac{\pi}{2}, \tfrac{\pi}{2}, 0, \pi \right) \otimes \hat{I} 
         = \begin{bmatrix} 1 & 0 \\ 0 & -1 \end{bmatrix} \otimes \hat{I}, \notag \\
\hat{O}_2 &= \hat{C}\left( \tfrac{\pi}{2}, \tfrac{\pi}{2}, 0, 0 \right) \otimes \hat{I} 
         = \begin{bmatrix} -1 & 0 \\ 0 & 1 \end{bmatrix} \otimes \hat{I}, \notag \\
\hat{O}_3 &= e^{i\pi} \left( \hat{C}(0, 0, 0, 0) \otimes \hat{I} \right) 
         = e^{i\pi} \left( \begin{bmatrix} 1 & 0 \\ 0 & 1 \end{bmatrix} \otimes \hat{I} \right) , \notag \\
\mathbb{I} &= \hat{C}(0, 0, 0, 0) \otimes \hat{I} 
          = \begin{bmatrix} 1 & 0 \\ 0 & 1 \end{bmatrix} \otimes \hat{I}.
\label{eq:operators_twoqubit_dj_usingauxiliary_qw}
\end{align}  
The schematic illustration of the quantum walk based oracle implementations for all two-bit boolean functions that are either constant or balanced  without using auxiliary qubit is shown in Fig~\ref{fig:dj_oracle_two_qubit_qw_without_aux}. From \eqref{Eq:HWP_JOnes_Matrix} and \eqref{eq:operators_twoqubit_dj_usingauxiliary_qw},  $\hat{O}_1 = \text{HWP}(0)$ and  $\hat{O}_2=\text{HWP}(\pi/2)$ so, the operations for the oracle implementation shown in Fig.~\ref{fig:dj_oracle_two_qubit_qw_without_aux} can be implemented in the photonic system a combination of phase shifters at an angle of $\pi$  and half-wave plates (HWPs) at angles $0$ and $\pi$  as shown in Fig.~\ref{fig:dj_oracle_two_qubit_qw_without_aux_on_ps} , where phase shifter at an angle of $\pi$ applies a phase shift of $e^{i\pi}$ to the mode where it is implemented i.e.,  $\hat{O}_3 = \text{PhaseShifter}(\pi)$. In the photonic implementation, a single photon performs a quantum walk where its polarization encodes the first qubit and its two paths (modes) encode the second qubit. The photon starts with  $|H\rangle$ polarization state in the 1st path ($|0\rangle$). Both  Hadamard transformation, given by equations~\eqref{eq:dj_superposiotion_state_without_auxiliary)} and~\eqref{eq:psi2_state_without_auxiliary}, are implemented using half-wave plates set at $\pi/8$ (HWP($\pi/8$))~\eqref{Eq:HWP_JOnes_Matrix} and 50:50 beam splitters~\eqref{Eq:BS_matrix}, as highlighted in (a) and (c) boxes in Fig.~\ref{fig:2dj_without_aux_ps_full}, respectively. The sequence of operations for the oracle implementation are depicted in Fig.~\ref{fig:dj_oracle_two_qubit_qw_without_aux_on_ps}.      The photonic circuit for the two-qubit Deutsch-Jozsa algorithm based on photonic quantum walk without using 
 auxiliary qubit for the function $f(x_1,x_2) = 0$ is shown in Fig.~\ref{fig:2dj_without_aux_ps_full}, where the (c) box represents the oracle. For the other functions listed in Table~\ref{tab:two_bit_functions}, this  box is replaced with the corresponding photonic circuit that implements the oracle for the given function, as shown in Fig.~\ref{fig:dj_oracle_two_qubit_qw_without_aux_on_ps}.

\section{Bernstein-Vazirani Algorithm} \label{Bernstein–Vazirani Algorithm}
The Bernstein-Vazirani algorithm is a quantum algorithm designed to determine a hidden binary string $s$ using a single query to an oracle. This problem is formally defined as follows: input is an oracle that realizes the function \( f: \{0,1\}^n \to \{0,1\} \) of the form \begin{align}\label{eq:bv_function}
f(x) &= x \cdot s = x_1 s_1 \oplus x_2 s_2 \oplus \dots \oplus x_n s_n,
\end{align} where $s$ is an unknown string of length $n$, the task is to determine $s$ efficiently.
Classically, determining $s$ requires querying the function at least $n$ times, each time using an input $x$ that isolates one bit of $s$. However, the Bernstein-Vazirani algorithm determines $s$ in a single query. The procedure is as follows:
\begin{itemize}
\item An $n$-qubit register is initialized in the state $|\phi_0\rangle=|0\rangle^{\otimes n}$, and a Hadamard transform $H^{\otimes n}$ is applied to create an equal superposition:
\begin{equation}\label{eq:bv_superposition_state}
|\phi_1\rangle = \frac{1}{\sqrt{2^n}} \sum_{x=0}^{2^n-1} |x\rangle
\end{equation} 
\item  The function $f(x)$ is embedded in an oracle operation $U_f$, which applies the transformation
\begin{equation}\label{eq:bv_Oracle_transformation1}
U_f |x\rangle = (-1)^{f(x)} |x\rangle,
\end{equation}
\begin{figure}[htbt]
    \centering
    \includegraphics[width=0.75\linewidth]{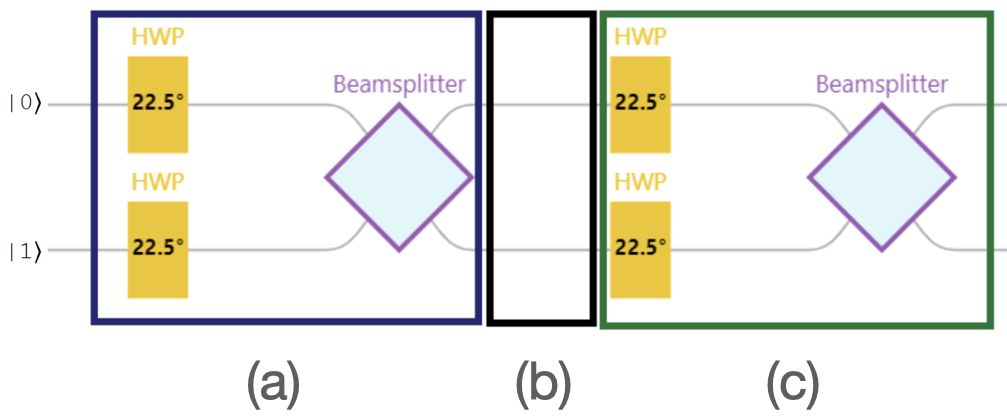}
    \caption{The photonic circuit for the photonic quantum walk based two-qubit Deutsch-Jozsa algorithm without using auxiliary qubit for the function $f(x_1,x_2)=0$.  In this setup, the polarization state and the two path degrees of freedom (\(|0\rangle, |1\rangle\)) correspond to the first and second qubits, respectively. The (a), (b), and (c) boxes denote the Hadamard (\(H\)) transformation on both qubits, the oracle, and another Hadamard (\(H\)) transformation on both qubits, respectively.
} \label{fig:2dj_without_aux_ps_full}
\end{figure}
where $f(x) = x \cdot s$, this results in
\begin{align}\label{eq:bv_phi2}
|\phi_2\rangle &= U_f |\phi_1\rangle = \frac{1}{\sqrt{2^n}} \sum_{x=0}^{2^n-1} (-1)^{x \cdot s}|x\rangle
\end{align} 
\item  Applying another Hadamard transformation to each qubit,
\begin{align}\label{eq:bv_phi3}
|\phi_3\rangle &= H^{\otimes n} |\phi_2\rangle =
\frac{1}{2^n} \sum_{y=0}^{2^n-1} \sum_{x=0}^{2^n-1} (-1)^{x \cdot s + x \cdot y} |y\rangle \notag \\ &= \frac{1}{2^n} \sum_{y=0}^{2^n-1} \sum_{x=0}^{2^n-1} (-1)^{x \cdot (s \oplus y) }|y\rangle = |s\rangle
\end{align}
\item Measuring the quantum register ( all $n$ qubits) determines the hidden string $s$ with certainty, as the sum evaluates to $1$ if $s = y $ i.e., $s \oplus y = 0$ , and $0$ otherwise.
\end{itemize}
The procedure described here does not utilize an auxiliary qubit; however, the same result can be achieved by incorporating an auxiliary qubit and applying the oracle transformation:  \begin{equation}\label{eq:bv_oracle_with_auxiliary)}
        U_f |x\rangle |y\rangle = |x\rangle |y \oplus f(x)\rangle.
\end{equation} The Boolean function \( f(x) = x \cdot s \)~\eqref{eq:bv_function} is a subset of constant and balanced functions. It is constant if \( s = 000\dots0 \) and balanced if \( s \neq 000\dots0 \), as it then outputs $0$ and $1$ an equal number of times over all \( 2^n \) inputs. 
Since the functions in the Bernstein-Vazirani algorithm form a subset of constant and balanced functions, the quantum walk scheme for the Bernstein-Vazirani algorithm, both with and without an auxiliary qubit, employs the same qubit encoding and Hadamard transformation as the corresponding quantum walk schemes for the Deutsch-Jozsa algorithm. Furthermore, for functions that belong to this subset, the oracles in the Bernstein-Vazirani algorithm~\eqref{eq:bv_oracle_with_auxiliary)},~\eqref{eq:bv_Oracle_transformation1} are identical to those in the Deutsch-Jozsa algorithm~\eqref{eq:dj_oracle_with_auxiliary)},~\eqref{eq:dj_oracle_without_auxiliary)}, utilizing the same set of quantum walk operations.  
For the Bernstein-Vazirani algorithm of two qubits, there are four possible hidden strings, as shown in Table~\ref{tab:two_bit_functions_bv} along with the functions associated with these strings. The oracle implementations in quantum walk schemes with and without an auxiliary qubit are shown in Fig.~\ref{fig:dj_oracle_two_qubit_qw} and Fig.~\ref{fig:dj_oracle_two_qubit_qw_without_aux}, while their implementations in a photonic system are presented in Fig.~\ref{fig:dj_oracle_two_qubit_qw_onps} and Fig.~\ref{fig:dj_oracle_two_qubit_qw_without_aux_on_ps}, where \( f(x_1, x_2) \) is defined as (\textit{i}) \(0\), (\textit{iii}) \(x_1\), (\textit{iv}) \(x_2\), and (\textit{vii}) \( x_1 \oplus x_2 \).  
\section{Discussion and Analysis}\label{Discussion and Analysis} 
\begin{figure}[htbt]\centering \includegraphics[width=0.9\linewidth]{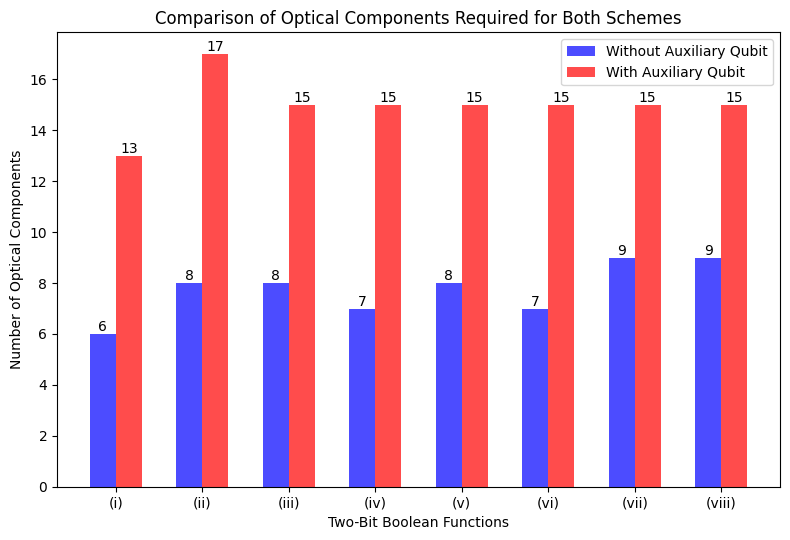}
    \caption{The plot comparing the optical components required for implementing the quantum walk schemes of the Deutsch-Jozsa and Bernstein-Vazirani algorithms, both with and without an auxiliary qubit, in a photonic system. The comparison includes all two-bit Boolean functions listed in Table~\ref{tab:two_bit_functions} for the Deutsch-Jozsa algorithm and the four two-bit boolean functions for the Bernstein-Vazirani algorithm defined as (\textit{i}) \(0\), (\textit{iii}) \(x_1\), (\textit{iv}) \(x_2\), and (\textit{vii}) \( x_1 \oplus x_2 \). The components essential for single-photon generation and measurement are not included in this comparison.} 
    \label{fig:plot}
\end{figure}

The quantum walk scheme for Deutsch-Jozsa algorithm presented in Section~\ref{Quantum walk scheme without using auxiliary qubit}, which does not utilize an auxiliary qubit, is more resource-efficient than the scheme described in Section~\ref{Quantum walk scheme using auxiliary qubit}, as as it eliminates the need for an additional auxiliary qubit and reduces the total gates required, thus lowering the overall circuit complexity. The plot in Fig.~\ref{fig:plot} presents a comparison of the \begin{table}[htbt]
    \centering 
    \caption{The four possible hidden strings and their corresponding functions in the two-bit Boolean function \( f(x) = x \cdot s \)~\eqref{eq:bv_function}.}
    \begin{tabular}{|c| c|}
        \hline
        Hidden String & Functions \( f(x_1,x_2)= \) \\  
        \hline
        00 &  \( 0 \) \\  
        01 & \( x_2 \) \\  
        10 & \( x_1 \) \\ 
         11& \(x_1 \oplus x_2  \) \\  
        \hline
    \end{tabular}
    \label{tab:two_bit_functions_bv}
\end{table} optical components required to implement both schemes in the photonic system for all the two-bit Boolean functions listed in Table~\ref{tab:two_bit_functions}. The component counts are determined based on the implementation of all gates, as depicted in Fig.~\ref{fig:dj_oracle_two_qubit_qw_onps}, Fig.~\ref{fig:2dj_aux_ps_full}, Fig.~\ref{fig:dj_oracle_two_qubit_qw_without_aux}, and Fig.~\ref{fig:2dj_without_aux_ps_full}. In this comparison, the mode-permuter is not considered a distinct physical component, as it can be naturally realized through path relabeling and the components essential for single-photon generation and measurement are also not included. For the quantum walk scheme for Deutsch-Jozsa algorithm using an auxiliary qubit, \( |00\rangle \) state of the working qubits corresponds to a photon present in the path state \( |00\rangle \). Therefore, only a single-photon detector at the first path is required.  For the quantum walk scheme for Deutsch-Jozsa algorithm without an auxiliary qubit, the state \( |00\rangle \) corresponds to a horizontally polarized (\( |H\rangle \)) photon in path $1$ (\( |0\rangle \)). Thus, both polarization and path state need to be measured, which can be achieved using a PBS or a polarizer along with a single-photon detector. The plot in Fig.~\ref{fig:plot} also presents a comparison of the two-qubit Bernstein-Vazirani algorithm under both schemes (with and without an auxiliary qubit) for the four functions, where \( f(x_1, x_2) \) is defined as (\textit{i}) \(0\), (\textit{iii}) \(x_1\), (\textit{iv}) \(x_2\), and (\textit{vii}) \( x_1 \oplus x_2 \). 
In the measurement stage of the Bernstein-Vazirani algorithm, both working qubits must be measured in the scheme with an auxiliary qubit using single-photon detectors at all paths. In contrast, in the scheme without an auxiliary qubit, all qubits must be measured, which can be achieved using a PBS or a polarizer along with single-photon detectors at all the paths to measure both polarization and path states. 

\section{Conclusion}\label{conclusion} 
In this paper, we have presented the scheme  for implementing the Deutsch-Jozsa and Bernstein-Vazirani algorithm using  single‑particle discrete‑time quantum walk, both with and without an auxiliary qubit. We have given a set of experimentally realizable walk operations that can implement the oracles for  the Deutsch-Jozsa and Bernstein-Vazirani algorithms along with the details to implement the operations in the photonic system. Detailed implementation of the oracles for two-qubit Deutsch-Jozsa and two-qubit Bernstein-Vazirani algorithm  using this set of operations is presented in the paper, and this set of walk operations is sufficient to implement the oracles for the $n$-qubit Deutsch-Jozsa and Bernstein-Vazirani algorithms, as an appropriate combination of these operations enables the realization of any oracle for both algorithms. Unlike the standard circuit model, where oracle implementations rely on CNOT gates, the quantum walk approach enables a more straightforward realization using a simple set of quantum walk operations.   A comprehensive analysis of the quantum walk schemes, with and without an auxiliary qubit in a photonic system, demonstrates that the scheme without an auxiliary qubit is more resource efficient, as it eliminates the need for an additional qubit and its associated quantum gates while also reducing the overall optical component count.  

\section*{Acknowledgment}\label{ack}
This work utilized \textit{Qniverse: A Unified Quantum Computing Platform}~\cite{Qniverse} for the construction of quantum circuits. This research was supported by the Ministry of Electronics and Information Technology (MeitY), Government of India, under the project titled “\textit{HPC-based Quantum Accelerators for Enabling Quantum Computing on Supercomputers}”, Grant No. 4(3)/2022-ITEA.

\end{document}